# Early Birds, Night Owls, and Tireless/Recurring Itinerants: An Exploratory Analysis of Extreme Transit Behaviors in Beijing, China

Ying Long, Beijing Institute of City Planning, China, longying1980@gmail.com
Xingjian Liu, The University of Hong Kong, China, xliu6@hku.hk
Jiangping Zhou*, Iowa State University, USA, zjp@iastate.edu
Yanwei Chai, Peking University, China, chyw@pku.edu.cn

* Corresponding author

**Abstract:** This paper seeks to understand extreme public transit riders in Beijing using both traditional household survey and emerging new data sources such as Smart Card Data (SCD). We focus on four types of extreme transit behaviors: public transit riders who (1) travel significantly earlier than average riders (the 'early birds'); (2) ride in unusual late hours (the 'night owls'); and (3) commute in excessively long distance (the 'tireless itinerants'); (4) travel over frequently in a day (the 'recurring itinerants'). SCD are used to identify the spatiotemporal patterns of these three extreme transit behaviors. In addition, household survey data are employed to supplement the socioeconomic background and provide a tentative profiling of extreme travelers. While the research findings are useful to guide urban governance and planning in Beijing, the methods developed in this paper can be applied to understand travel patterns elsewhere.

**Acknowledgments:** The authors would like to acknowledge the financial support of the National Natural Science Foundation of China (No.51408039) and the 12[th] Five-year National Science Supported Planning Project of China (2012BAJ05B04). Any errors and inadequacies of the paper remain solely the responsibility of the authors.

# 1 Introduction

Extreme events and mechanisms behind them often capture our attentions: the most dominant city of a nation, the most depressed city of a region, the most popular gateway city among immigrants, etc. and associated underlying mechanisms such as what contributes to the extreme and how they are related to other phenomena in between. In the past decade or so, against the backdrops of the global financial crisis, increased numbers of the unemployed, self-employed and part-timers, rise of telecommuters and low-paying jobs relocated to cheaper places inside or outside a region/country, extreme travelers have received increasing attention from both academia and mass media in recent years. This strand of research concerns travelers making unusually long, early, late, or other extreme daily trips, especially long commuting trips, which account for a significant portion of trips that residents make daily (Barr, et al., 2010; Gregor, 2013; Jones, 2012; Landsman, 2013; Marion and Horner, 2007; Moss and Qing, 2012; Rapino and Fields, 2013; U.S. Census, 2005).

While scholarly work on extreme travelers has been largely underdeveloped for Chinese cities, we are particularly interested in extreme travelers using public transit. For one thing, Chinese cities have historically relied on public transit. For another, the Chinese government has sought public transit as a major remedy to deal with congestions, pollutions, and other issues related to rising car ownership.

To date, extreme traveler analyses have mostly drawn upon traditional data such as travel diaries and household surveys. More recently, emerging big data sources such as transit smart card data have not been utilized to investigate the phenomenon. Transit smart card data record rich information about individual trips (e.g., origin, destination, and time length) and thus seem to be useful supplementary sources for understanding travel behaviors. For example, since the 1990s, use of smart cards has become prevalent in more and more cities owing to the development of the Internet and the increased complexity of mobile communication technologies (Blythe, 2004). Still, Intelligent Transportation Systems (ITS) that incorporate smartcard-automated fare systems had been in place in over 100 Chinese cities as of 2007 (Zhou, 2007). The combination of these conventional and emerging data sources offers new opportunities to deepen our understandings of cities and related routine activities such as traveling and commuting (Batty, 2012).

While most literature on extreme travelers focuses on excessively long trips (the 'tireless itinerants'), we extend the tireless itinerants as those who travel over frequently and meanwhile concern two additional types of extreme transit behaviors: public transit riders who (1) travel significantly earlier than average riders (the 'early birds') during weekdays; and (2) ride in unusual late hours (the 'night owls') during weekdays. More specifically, we seek to identify these extreme commuters in Beijing, characterize their spatiotemporal trajectories, profile their socioeconomic backgrounds, and tentatively generalize the dynamics behind extreme travel behaviors. This study would not only provide insights into travel patterns in in a representative populous Chinese city, but also explores a generic analytical framework that combines traditional and emerging data sources to understand urban travel behaviors.

# 2 Data

Our study area – Beijing Metropolitan Area (BMA), China – covers an area of 16,410 km$^2$ and has a population of more than 20 million in 2010. Gaining momentum from China's recent economic success, Beijing, as the capital city, is becoming one of the world's most populous and fastest growing metropolis. (See Yang et al. (2011) for more background information about Beijing).

Beijing's public transit system consists mainly of buses and subway lines. As of 2010, there were 184 kilometers of subway lines (excluding the airport express rail) in Beijing. In 2011 alone, the bus system delivered 4.9 billion rides and travelled 1.7 billion kilometers. Thanks to the continuous expansion of subway lines as well as subsidies for public transit, the share of subway and bus trips among all commutes in BMA has risen steadily, reaching 38.9% in 2010, making Beijing the largest public transit system in terms of daily ridership.

## 2.1 Beijing bus/metro smart card data in 2010

Since 2005, over 90% of bus/metro riders in Beijing use anonymous smart cards to pay their fare. This high rate of smart card usage among bus/metro riders is largely due to governmental subsidies for public transit. For example, riders using smart card data enjoy 60% discounts on any routes in the local bus system, and the discount rate for students is at 80%. Smart cards also enable cardholders to pay for other services such as taxi, electricity, and sewage that are offered by the local government or government linked companies.

When cardholders use their smart cards to pay for bus/metro services, card readers installed on the bus and in the station automatically record information about trip origin and/or destination stops/stations, boarding and/or alighting time, smart card numbers, as well as the card type (e.g. student cards versus regular cards).

Information collected for bus trips needs some further elaboration. There exist types of bus fares. The first fixed fare is for short routes most of which are within the fifth ring road, while the second distance based fare is associated with long routes. For the first type, 0.4 Chinese Yuan (CNY; approximately 0.06 USD) is charged for individual bus rides, and the corresponding SCD contains only the departure time and stop ID and no arrival time or stop ID. Cardholders' spatiotemporal information is incomplete for this kind of route. Our strategy for tackling this incomplete information issue of fixed-fare records have been elaborated in Long and Thill (2013), and the strategy proposed by Ma et al (2012) could be used in future for solving the incomplete information of Beijing bus SCD. For the distance-based fare type, the amount charged depends on the route ID and trip distance, and the SCD contains full information. Both types of bus SCD records are used in this paper for analyzing extreme bus/metro riders and profiling their mobility patterns. Still, one subway ride costs 2 CNY, regardless of the trip distance and time, and the subway records have complete spatiotemporal information.

We collected one-week bus/metro SCD in 2010 (5-11 April, Monday to Sunday) from Beijing Municipal Administration & Communications Card Co. (BMAC). The dataset records 97.9 million trips (59.3m fixed fare bus trips, 23.4m distance based fare bus trips, and 15.2 metro journeys) for 10.5 million cardholders. SCD have been geocoded using bus/metro routes and the locations of stops/stations (Figure 1). Only SCD records in weekdays (5-9 April) are used in the subsequent analysis, and then there are 9.4 million active cardholders.

## 2.2 Transit system and traffic analysis zones (TAZs) of Beijing

As already implied, GIS data about transit lines (locations of routes and stops for buses, as well as lines and stations for the subway) are essential for geocoding and mapping SCD. In 2010, there were 1928 bus routes (Figure 1a) and 21,372 bus stops (Figure 1b) in the BMA. Note that a bus route could have two directions, e.g., the bus No. 113 has two routes, one from Dabeiyao to Qijiahuozi and the other from Qijiahuozi to Dabeiyao. These routes are counted separately in this paper. The average distance between consecutive bus stops is around 231 m. In 2010, there are 9 subway lines, including the airport express line, and 147 subway stations associated with (Figure 1c). We use Beijing TAZ data to aggregate the analytical results for better visualization. In total, 1,911 TAZs are defined (Figure 1d) according to the administrative boundaries, main roads, and the planning layout in the BMA.

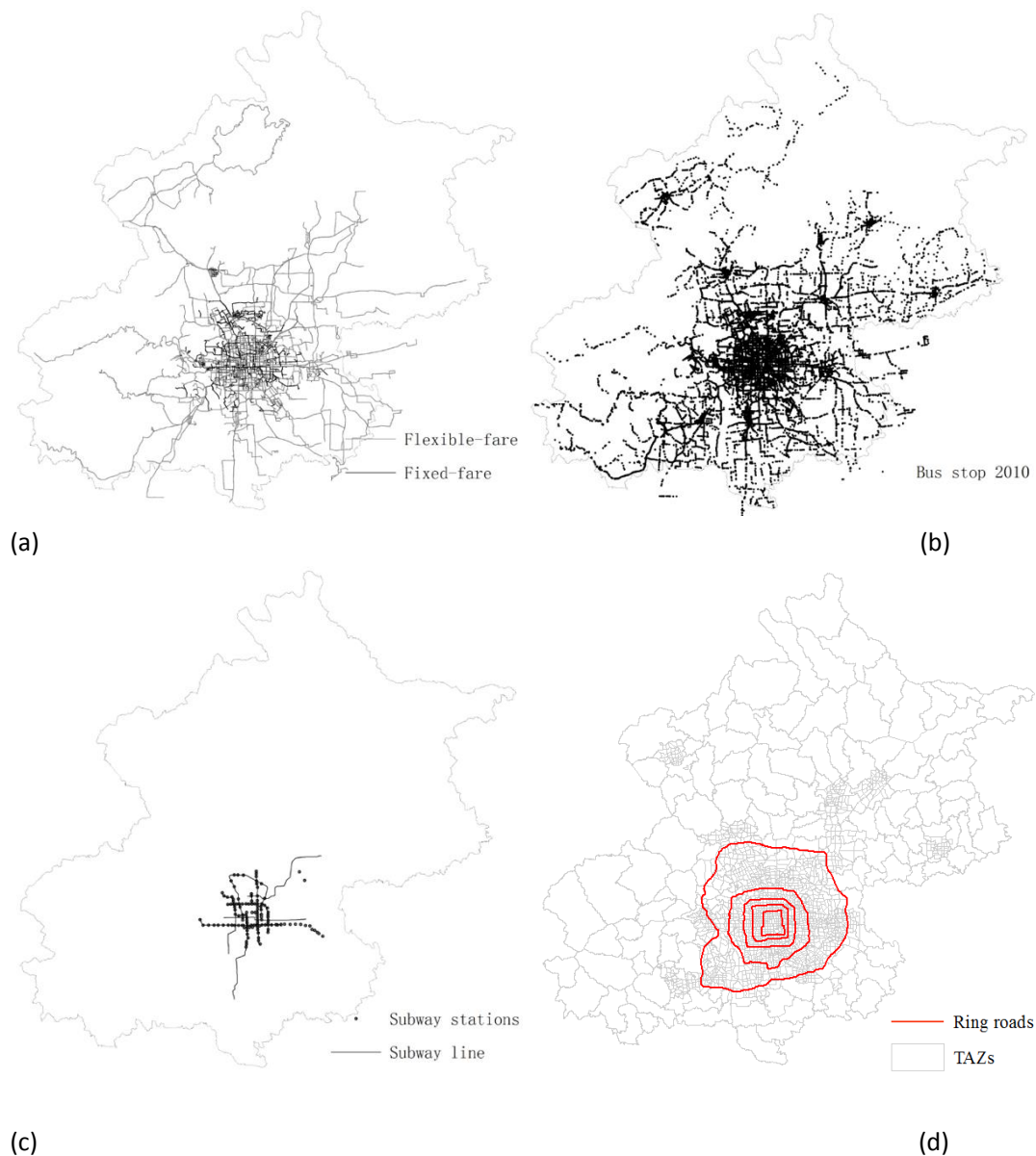

(a) (b)

(c) (d)

Figure 1 Bus routes in 2010 (a), bus stops in 2010 (b), subway lines and stations in 2010 (c), and traffic analysis zones in 2010 (TAZs) (d) of the BMA. Note: All maps are from the Beijing Institute of City Planning. Some bus routes and stops are outside the BMA, as shown in a-d since some residents live outside the BMA and in adjacent towns in Hebei province.

### 2.3 Household travel survey of Beijing in 2010

The 2010 Beijing Household Travel Survey (the 2010 survey hereafter) is used to profile individual extreme travelers, or more specifically, extreme transit riders in the city. The 2010 survey represents the conventional approach to understanding transit behaviors. This survey adopts a multistage sampling strategy with a targeted 1% sampling rate. 1,085 out of 1,911 TAZs in the entire BMA are featured in this survey, with the sparsely populated TAZs excluded. In each TAZ, 10 to 50 households are selected to take a face-to-face interview. The final sample consists of 46,900 households (116,142 residents) in the BMA. The 2010 survey provides one-day travel diaries of all respondents, which gives travel time for each employee (distance unavailable except the TAZs of Origin-Destination). The travel mode is identified as the mode of the trip segment with the longest duration. We choose the mode of the highest mobility when there are two or more segments that have the same, longest duration. The 2010 survey presents also household information including household structure, income, and residential location at the TAZ level, as well as personal information including gender, age, occupation, industry of employment, etc. Like SCD, only records surveyed in weekdays are pertained in the subsequent analysis.

## 3 Methods

Our analytical approach can be outlined as follows: Firstly, we set working definitions for different types of extreme travel behaviors and consequently identify extreme travelers from the SCD dataset. Secondly, we characterize the spatiotemporal trajectories of extreme travelers. Lastly, we supplement the SCD data with the household survey to profile the socioeconomic conditions of identified extreme travelers.

### 3.1 Defining and identifying extreme travelers from the SCD

We define four types of extreme travelers (tireless itinerants classified into two types in the following discussion) in Beijing according to their transit behaviors in weekdays (Table 1). These working definitions draw upon the 2010 survey, existing literature, as well as researchers' own experiences of living in Beijing. For instance, the regular working hour starts on 8:30 or 9:00 am in Beijing, and therefore boarding public transit before 6:00 am would be considered as unusually early.

**Table 1 Working definitions of extreme travelers**

| Type | Definition |
|---|---|
| Early Birds (EBs) | First trip < 6AM, more than two days in a week (60% of weekdays) |
| Night Owls (NOs) | last trip (boarding time) > 10PM, more than two days in a week (60% weekdays) |
| Tireless Itinerants (TIs) | >= one and a half hours commuting, more than two days in a week |
| Recurring Itinerants (RIs) | >= 30 trips in weekdays of a week (>= 6 trips per day) |

### 3.2 Mobility patterns of extreme travelers based on SCD

To analyze the mobility pattern of identified extreme travelers, we construct commuting journeys based on commuters' job and home locations, which are in turn inferred using the following procedure:
- The card type is not a student card;
- $D_j>=6$ hours, where $D_j$ is the duration that a cardholder stays at place j, which is associated with all bus stops within 500 meters of one another. Note that the benchmark of 6 hours are set based on the analysis on fulltime job in the 2010 survey;
- j <>1, which means that j is not the first place in a weekday that the server records;
- The place where a cardholder visited most frequently in five weekdays will be defined as the final workplace of the cardholder in this study.

Similarly, we deduced from the data queries that a place would be a cardholder's housing place if the data meet these conditions:
- The card type is not a student card;
- The place where a cardholder's first boarding bus stop/subway station trip in a day most frequently in five weekdays will be defined as the final housing place of the cardholder in this study.

Second, commuting trips were then identified based on the identified job and housing places of a cardholder. More details about identifying housing and job locations are available in Long and Thill (2013). In addition, we plotted three typical extreme travelers for each type to gain more knowledge.

### 3.3 Probing socioeconomic characteristics and the travel purpose for identified extreme travelers by using the 2010 survey

Big data do not inform us about individual extreme travelers and reasons behind their traveling behaviors. We probed into these issues by using the 2010 survey samples mentioned above. The counterpart of each identified extreme travelers from the SCD was extracted from the survey with the same rules we mined SCD for extreme travelers. Using the 2010 survey, we analyze the first bus/metro trip of EBs, the last bus/metro trip of NOs, the bus/metro commuting trip of TIs, all bus/metro trips of RIs, and all trips of surveyed residents (Average Beijingers, ABs).

## 4 Results

Based on our working definitions of extreme travelers, 188.9 thousand or 2.0% of all active cardholders in weekdays are associated with one or more extreme travel behaviors (the percentage is consistent with the findings of Rapino and Fields (2013) in the U.S.). 14.2k of these extreme travelers use student smart cards. We also note that 4,890 cardholders fit two types of extreme travel behaviors, and 40 are three types. We find that there are more RIs and NOs than the other two types of extreme travelers, and TIs are the fewest in 2010.

**Table 2 The inventory of four types of identified extreme travelers**

| Type | # cardholders (*1000) | # students cards (*1000) |
|---|---|---|
| EBs | 42.8 | 2.8 |
| NOs | 70.6 | 4.9 |
| TIs | 6.7 | 0 |
| RIs | 73.7 | 6.8 |

The mobility patterns of four types of extreme travelers are summarized in Table 3. For housing, there are a significant percentage of TIs residing in the three well-known suburb residential areas that were built to accommodate local relocated residents: Tongzhou, Huilongguan, and Tiantongyuan. In addition to the three areas, EBs distribute in the surrounding areas of central city, RIs cluster in the SE of central city (the area within the fifth ring road of Beijing), and NOs concentrate within the second and fourth ring road. For jobs, most of extreme travelers' job locations are within the fifth ring road and are in the northern part of the city, which is more developed comparing with the southern part. Many RIs work at the Yizhuang Industrial Park. According to the survey, most of them reside in the central city of Beijing. A notable number of TIs work at the Shangdi Information Technology cluster and around Tiantongyuan. For commuting trips, the popular destinations of EBs' commuting was to Xizhimen area. Many TIs commute a long distance from Tongzhou to the northern part of the central city. Most TIs commute from outside the central city to the central city. Many RIs commute a long distance, although only 4.0% of all identified RIs have an identified commuting trip. For typical trips of three extreme travelers in Table 2, there are only commuting trips for EBs and NOs, and not all EBs and NOs commute a long distance. In addition, RIs visit several places and travel a long distance in a day. In addition, we analyze all identified extreme travelers' trips, EBs, NOs and TIs have very few non-commuting trips (27.4%, 25.3% and 36.8% respectively), indicating that they are busy with working activities.

**Table 3 Mobility patterns of four types of extreme travelers**

| Extreme travelers | Kernel density of housing | Kernel density of jobs | Commuting trips | Most visible trips |
|---|---|---|---|---|
| EBs | 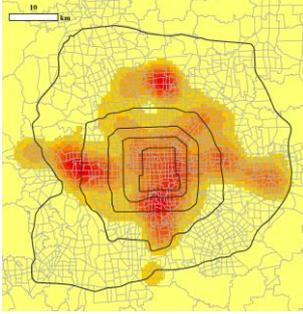 (10.3 k) | 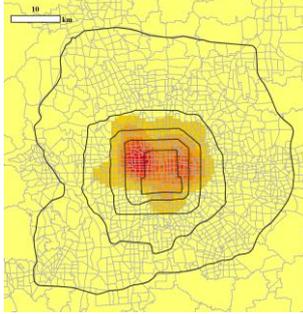 (9.4 k) | 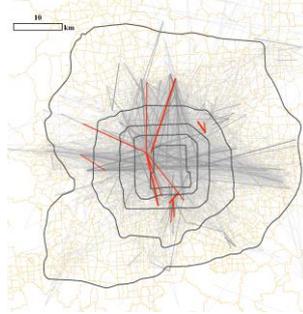 (4.9 k) | 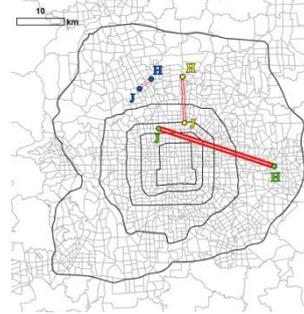 |
| NOs | 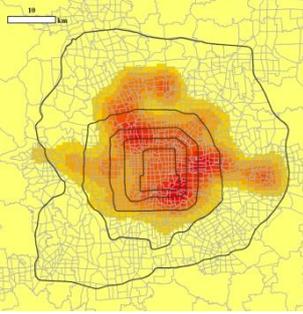 (31.6 k) | 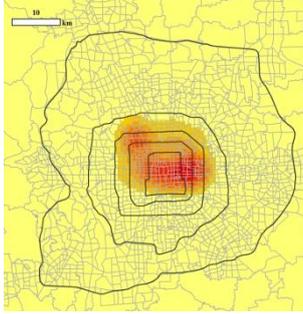 (25.0 k) | 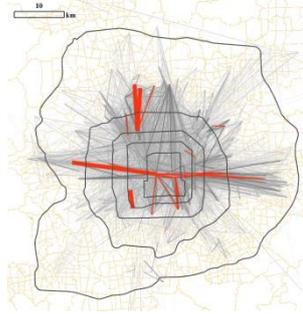 (17.5 k) | 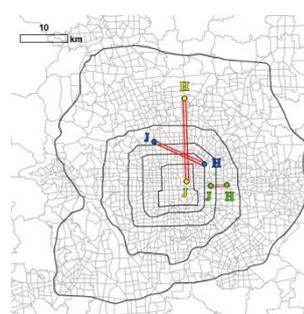 |
| TIs | 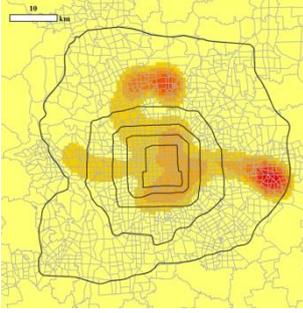 (6.7 k) | 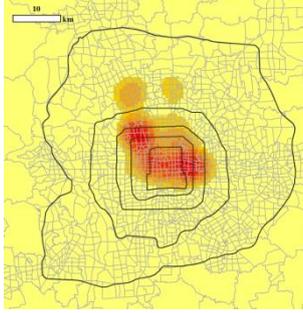 (6.7 k) | 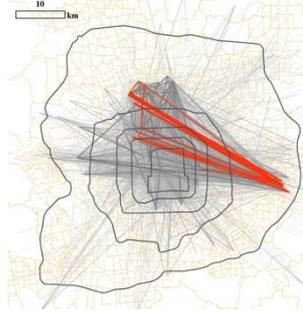 (6.7 k) | 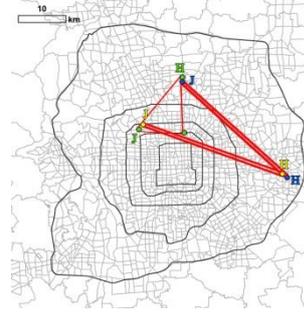 |
| RIs | 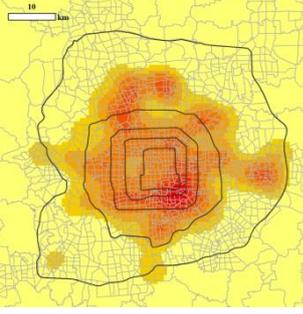 (25.4 k) | 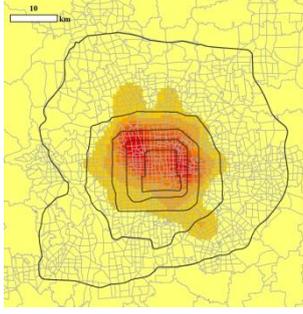 (7.8 k) | 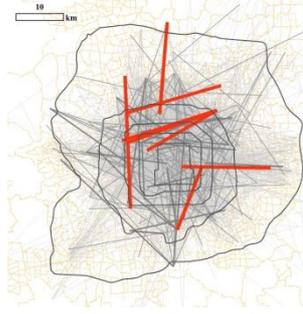 (2.7 k) | 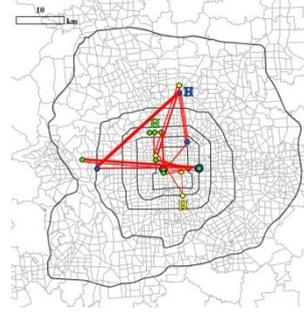 |
| Legend | 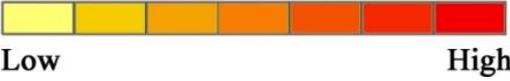 | | 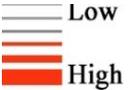 | |

Note that not all cardholders have a housing or job place, and a commuting trip. Commuting maps are prepared by using the head/tail breaks approach proposed by Jiang (2013) for data with a heavy-tailed distribution. In the maps of typical trips, "H" is for a housing place, and "J" is for a job place. Numbers in brackets are the total count of extreme travelers with corresponding information. The base layer is the TAZs in 2010. Ring roads in each map correspond to those in Figure 1(d).

Overall 1,569 or 7.2% extreme travelers are identified from all 21,771 travelers with at least one bus/metro trip in weekdays as documented in the 2010 survey. There are 676 EBs, 236 NOs, 627 TIs, and 100 RIs (70 travelers fit two types of extreme travel behaviors). As shown in Figure 2, we found that 60.2%, 11.8% and 10.9% EBs' first trips are to workplace, school and recreation place, respectively. These are significantly different from those of ABs (21.9%, 3.9% and 6.4%). Most of NOs' last trips are to home (96.2%), and 2.1% are to workplace at night. Almost one third of RIs trips are for dining (31.2%), and a considerable amount of trips are for pick-up and drop-off as well as business errands (all greatly than ABs).

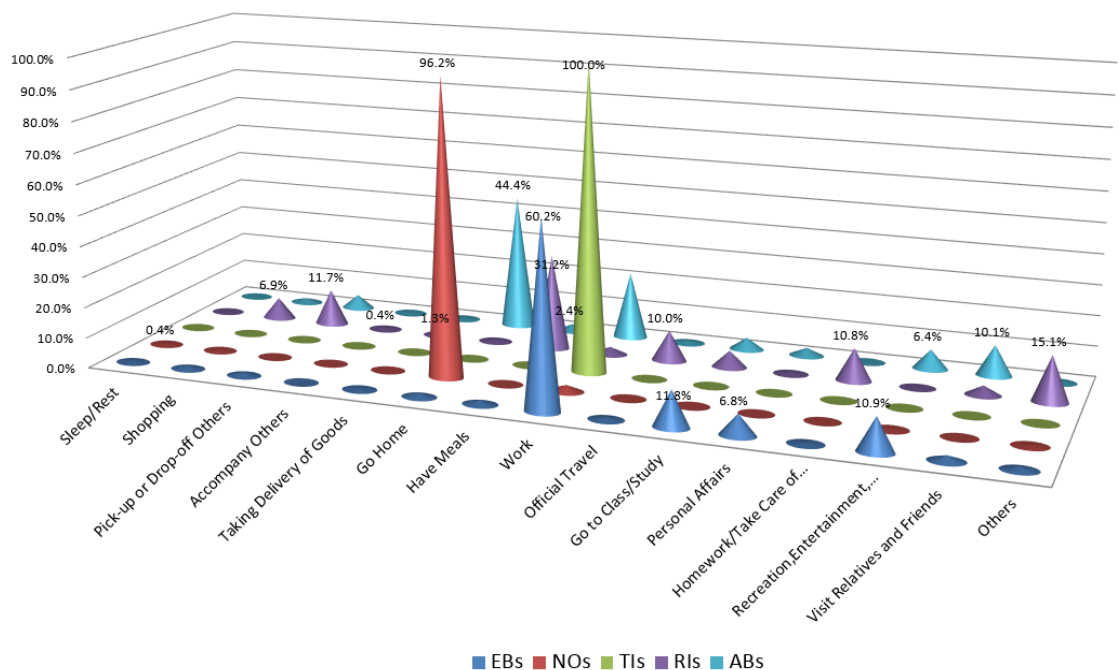

**Figure 2 Travel purposes of each identified extreme traveler from the 2010 survey**

We further analyzed socioeconomic characteristics of different extreme travelers and their households. Table 3 summarizes our main findings.

**Table 4 Selected socioeconomic characteristics of extreme travelers**

| ID | Extreme travelers | EBs (676) | NOs (236) | TIs (627) | RIs (100) | ABs (116,142) |
|---|---|---|---|---|---|---|
| 1 | % annual household income>=100 k CNY | 4.9 | 4.2 | 6.7 | 5.0 | 7.4 |
| 2 | % renting house | 11.0 | 17.8 | 20.4 | 16.0 | 16.1 |
| 3 | # average household car ownership | 0.22 | 0.21 | 0.25 | 0.22 | 0.31 |
| 4 | % higher education (undergraduate and graduate) | 14.2 | 18.2 | 33.5 | 25.0 | 21.1 |
| 5 | % Beijing Hukou | 87.0 | 82.2 | 74.8 | 83.0 | 82.4 |
| 6 | % public-sector employees | 13.5 | 7.6 | 15.8 | 7.0 | 10.4 |
| 7 | % fulltime workers | 60.9 | 84.7 | 94.4 | 42.0 | 45.9 |
| 7 | % fulltime students | 12.7 | 2.1 | 1.3 | 1.0 | 7.3 |
| 7 | % retirees | 20.9 | 5.9 | 0.8 | 38.0 | 29.1 |

Note that numbers in brackets are the total count of extreme travelers.

(1) Classified households with 100 k CNY as high-income ones, we found that all households consisting of extreme travelers have a lower ratio as compared to AB households.
(2) There are more residents who rent housing in Beijing in the groups of NOs and TIs, comparing with ABs.
(3) All four types of extreme travelers on average owned fewer cars than ABs.
(4) Tis and RIs have more percentage of higher education in contrast to ABs.
(5) TIs had the lowest percentage of local Beijing residences (Hukou) among all extreme travelers. It should be mentioned that there are around 40% residents in Beijing without Hukou in 2014. The 2010 survey is with bias in terms of this attribute.
(6) Fewer public-sector employees, who include civil servants and employees of public institutions are NOs and RIs. In Beijing, public-sector employees tend to have more stable jobs and higher income. More third-sector workers are NOs and TIs, and more employees in private companies are TIs. In addition, very few NOs are teachers and medical staffs. There are no soldiers and policemen in NOs, TIs and RIs.
(7) Regarding social status of all extreme travelers, 60.9% EBs are fulltime workers, 20.9% retirees, followed by 12.7% fulltime students. Most of NOs and TIs are fulltime workers. There are also a significant number of retirees in NOs (5.9%). Surprisingly, of all RIs, 38.0% are retirees while 42.0% are fulltime workers. There are also 12.0% of all RIs are jobless.

We are able to draw pictures for all four types with the analysis using the 2010 survey. For instance, most EBs are working full time in the private tertiary sector, fulfilling jobs that are less well-paid and require less educational attainment. Most of Tis are those who are also busy with working, with a higher percentage for renting house, well educated, with a lower percentage for holding Beijing Hukou, and working for private sectors.

# 5 Conclusions

In this paper, we first extended existing definition on extreme travelers and proposed four types of extreme transit riders including Early Birds, Night Owls, Tireless Itinerants and Recurring Itinerants. We then identified and profiled each type of extreme transit riders using both big and traditional data. Smart card data are used to visualize the overall spatiotemporal patterns of 188.9 thousand extreme travelers (2% of all active cardholders): where they reside and work, and what their mobility patterns are. In addition, we located 1,568 extreme travelers in the traditional household survey data as the counterparts identified from the SCD, and probed the travel purpose and socioeconomic attributes of extreme travelers and their households. To the best of our knowledge, this is a first attempt at understanding the full picture of extreme travelers with SCD for Chinese cities.

The overall contribution of this paper lies in three aspects: (1) we develop a new analytical framework to generate useful information about four types of extreme public transit riders from increasingly available big data such as smart card data; (2) we characterize the overall spatial trajectories as well as socioeconomic conditions of extreme transit riders; (3) the paper demonstrates how conventional and emerging data sources can be combined fruitfully to understand urban dynamics. More specifically, Long and Thill (2013) use rules identified from traditional surveys to guide the query of SCD and derive bus travelers' spatial information of residences and workplaces. In this paper, we further show how big data can help disclose overall patterns of all and extreme transit riders, which traditional data are unable to or are not good at. In addition to the above, we think our specific findings and workflows can be useful to local decision-makers. We found, for instance, the spatial distribution of RI, NO and EBs' homes and residences are similar and RIs and EBs both tend to have their workplaces inside the fifth ring road. We also found how lots of TIs cluster near the east border of the city while work in inner city. These finding can help local decision-makers envision and design better jobs-housing policies.

Future work could be expanded in at least three directions. First, only one week SCD in 2010 are employed in this paper. We would like to use the SCD in multiple years to gain more knowledge on the longer-term dynamics of extreme transit riders. In this regards, Batty (2013), rightly pointed out "[T]he power of big data is that if collected for long enough then the longer term will emerge from the short term. At the moment these data are about what happens in the short term, but over ten years or longer we will have a unique focus on the longer term – in fact we will have a snapshot of urban dynamics which is unprecedented." Second, the big data and small data we used to supplement each other could represent different aspects of the same universe. For instance, senior riders are not included in SCD and most respondents in the household survey are local residents with a Beijing Hukou. We therefore may have to design separate surveys to better understand extreme transit riders. Otherwise, our results can be biased. In the long run, if we want to avoid related biases, we need better strategies to collect and use information based on SCD, as recommended by Pelletier et al. (2011). Third, our analysis is restricted to extreme travel behaviors of public transit riders and do not cover travelers using other modes of transportation (e.g., car). Comparing extreme travel behaviors of different modes could shed more light on underlying mechanisms regarding what contributes to the extreme and how the extreme could have burden or benefit the travelers.